\documentclass{article}
\usepackage{epsfig,cite}
\usepackage{amssymb,amsmath}
\usepackage{times}

\author{Claudio Dappiaggi\footnote{claudio.dappiaggi@pv.infn.it}\\ 
Dipartimento di Fisica Nucleare e Teorica \& Sezione INFN di Pavia\\
I-27100 Via Bassi, 6  Pavia (Italy)}

\title{Can we implement the holographic principle in asymptotically flat
spacetimes?}

\begin{document}
\maketitle

\begin{abstract}
\noindent We discuss some recent results in the quest to implement the holographic principle in 
asymptotically flat spacetimes. In particular we introduce the key ingredients of 
the candidate dual theory which lives at null infinity and it is invariant under
the asymptotic symmetry group of this class of spacetimes.
\end{abstract}

\section{Introduction}
The last few decades witnessed the rise of several and often antithetical methods aimed at the 
quantization of general relativity. Unfortunately, up to now, none of them has been recognized as fully
satisfactory and the quest to derive a quantum version of Einstein theory still rages on. Beside the non
renormalizability of Einstein-Hilbert Lagrangian in a perturbative scheme, one 
of the main obstruction to the success
of the above programme lies in the existence of peculiar objects such as black 
holes which are the source
of extreme gravitational fields. Already at a classical level, their behaviour 
drastically differs 
from the usual physical systems commonly studied at low energies; to support such 
assertion, one simply needs to recognize that it is possible to associate to a 
black hole - considered as a dynamical system -
three ``evolution'' laws which are directly intertwined with the laws of thermodynamics (see \cite{Wald} 
and references therein for a detailed analysis). In particular, to a black hole,
one associates an entropy function related to its geometrical data by means of the widely-known 
Bekenstein formula\footnote{Unless stated otherwise, in this paper we assume 
$c=\hbar =G=1$}
\begin{equation}\label{1}
S=\frac{A}{4},
\end{equation}
where $A$ is the area of the event horizon. This relation shows a drastic departure from the 
behaviour of any other classical systems where entropy is proportional to the volume of the considered 
region of spacetime. Starting from these premises, 't Hooft remarked that a further striking consequence 
of (\ref{1}) arises if one considers $S$ as a (indirect) measure of the number of degrees of freedom accessible to 
a physical theory formulated inside the black hole itself; the key suggestion is that the 
proportionality of $S$ with the area could be interpreted as a signal that all
the phsycal information stored inside the black hole could be encoded by means of
a suitable second theory intrinsically formulated on 
the event horizon and with a density of data not exceeding the Planck density \cite{Hooft}. This conjecture 
goes under the name of \emph{holographic principle} and, in the last decade, it has been the driving 
principle behind several new achievements in quantum field theory. 

For our purposes it is imperative to mention only two key steps which have 
improved the original 't Hooft proposal; the
first goes under the name of ``covariant entropy conjecture'' which 
generalizes a Bekenstein-like formula 
to a wider class of spacetime regions \cite{Bousso}. In detail, consider in a $4$-dimensional Lorentzian 
manifold $(M,g_{ab})$ a spacelike 2-surface $\Sigma$ and the light rays originating from $\Sigma$ whose
congruences form several light-sheets. To each of these sheets, it is possible to associate the Raychauduri 
equation which governs the law of evolution for the area $A^\prime$ of $\partial\Sigma$ along the congruences;
the conjecture proposed by Bousso states that, whenever $A^\prime$ is
monotonically non-increasing and the sheet $L_\Sigma$
under consideration terminates on the boundary of the spacetime (or on a caustic), the entropy $S(L)$ 
of the gravitational and the matter fields evolving inside this region of the manifold is bounded by   
$S(L)\leq\frac{A^\prime}{4}$. 

Despite the above conjecture has been demonstrated under some specific and peculiar conditions (see \cite{Bousso2} 
and references therein), an almost straightforward consequence of Bousso 
conjecture is the chance to extend the holographic principle to any region where
the above bound holds: all the information of a theory living inside $L$ can be
encoded on the boundary of the sheet with a density of date not exceeding the
Planck density. Furthermore the above argument can be reversed
and it is possible to ask ourselves if, in a generic but fixed manifold (usually
called the ``bulk''), it exists a codimension 1 submanifold, also called
``screen'' (usually but not necessary the
boundary), which plays the role of the light sheet $L$ i.e.
all the information of a theory living in the whole spacetime can be encoded on 
it. This is a key consequence of the covariant entropy conjecture since it 
allows us to identify, by means only of a geometrical and 
classical construction, certain hypersurfaces in the spacetime 
as natural candidates where to construct the ``holographic theory''. 

Nonetheless 't Hooft original conjecture lacks any explicit mean to 
implement the holographic principle even when it is identified a screen 
where to encode the bulk data. Thus one has to look for a way to formulate the 
theory on the screen itself case-by-case.
A concrete realization of this paradigm has been proposed a few years ago and it
goes under the name of AdS/CFT correspondence \cite{Aharony} which states the
existence of a one to one correspondence between
a superstring theory of type IIB living on $AdS_5\times S^5$ and a $SU(N)$ super Yang-Mills field theory
living on the boundary of $AdS_5$. Although such conjecture is widely accepted and successfully tested
within different context, it strictly requires that an (asymptotically) AdS boundary condition is imposed
both on the underlying manifold and on the physical fields. It is thus natural to ask ourselves whether
such a conjecture and, in general, the holographic principle can be implemented whenever we consider a 
different class of spacetimes. In particular, in this paper we discuss some 
recent results on the scenario with a vanishing cosmological constant.

\section{Asymptotically flat spacetimes and the BMS group}
Starting from the premises discussed in the previous section, the first step to implement the holographic 
principle in an asymptotically flat spacetime consists on choosing a suitable codimension 1 submanifold
where to encode the data from the bulk theory. According to the covariant entropy conjecture \cite{Bousso}, 
the most natural candidate is future or past null infinity and, due to its importance, we
now review the key details of its construction though we refer to \cite{Wald} and section 2 in \cite{Dappiaggi3}
for a more exhaustive analysis. 

A manifold $(\hat{M},\hat{g}_{ab})$ is called asymptotically flat at null infinity if 
it exists a second manifold $(M,g_{ab})$, a positive scalar function $\Omega$ over $M$ (usually called 
``compactification factor'') and an embedding $i$ of $\hat{M}$ in $M$ such that 
\begin{itemize}
\item $\Im^+\cup i_0\cup\Im^-=\partial i(\hat{M})$ where $\Im^\pm$ are respectively future and null
infinity and where $i_0$ is spatial infinity,
\item $\hat{g}_{ab}=\Omega^2 g_{ab}$ 
on $i(\hat{M})$, $\Omega$ being everywhere smooth except at most $i_0$ where it is at least twice 
differentiable,
\item $\Omega=0$, but $d\Omega\neq 0$ point wisely on $\Im^\pm$.
\end{itemize}
According to the above construction both $\Im^+$ and $\Im^-$ are differentiable null manifolds topologically
equivalent to $\mathbb{R}\times S^2$; consequently, given a fixed asymptotically flat spacetime 
$(\hat{M},\hat{g}_{ab})$ and a fixed compactification factor $\Omega$, its boundary structure at 
(future or past) null infinity is \emph{intrinsically} characterized
by the following data: $\Im^\pm$, the restriction of the metric 
$h_{ab}=g_{ab}\left|_{\Im^\pm}\right.$ and the vector $n^a=g^{ab}\nabla_b\Omega$.
Furthermore this triple is also \emph{universal} i.e., given any two asymptotically flat manifolds, say
$(\hat{M}_1,\hat{g}_{1ab})$ and $(\hat{M}_2,\hat{g}_{2ab})$, let us associate to each of them an 
arbitrary triple defining the (future) null boundary structure, say $(\Im_1^+,h_{1ab},n_2^a)$ and 
$(\Im_2^+,h_{2ab},n_2^a)$. Then it always exists a diffeomorphism 
\begin{equation}\label{diffeo}
\gamma:\Im_1^+\to\Im_2^+
\end{equation}
such that $\gamma^*h_{2ab}=h_{1ab}$ and $n_2^a=\gamma_*n_1^a$ which implies that the boundary structure is the
same independently from the chosen bulk manifold and, more important, from the specific field theory
living on it. For this reason, $\Im^+$ and $\Im^-$ do represent the natural and more general framework 
where to construct an holographic theory for an arbitrary asymptotically flat spacetime.

The set of all maps, defined as in (\ref{diffeo}), constitutes the diffeomorphism group of 
the (future or past) boundary data  and its explicit representation can be provided once a suitable coordinate
frame on $\Im$ is chosen. The answer to this query goes under the name of ``Bondi frame'' $(u,\Omega,\theta,\varphi)$
where, besides the natural $(\theta,\varphi)$-coordinates over $S^2$ and the
conformal factor $\Omega$,
we have introduced $u$, the affine parameter along the null geodesics over $\Im$, i.e.
the integral curves of the vector $n^a$ which turns out to be complete. This specific choice
of coordinates can be restricted to $\Im$ simply remembering that, by construction, $\Omega\left|_\Im\right.=0$
and, thus, the above set of diffeomorphisms becomes the following set of 
transformations ($z=ctg\theta e^{i\varphi}$):
\begin{eqnarray}
u\longrightarrow u^\prime=K_\Lambda (z,\bar{z})\left[u+\alpha(z,\bar{z})\right],\label{3}\\
z\longrightarrow z^\prime=\frac{az+b}{cz+d}\label{4},\\
\bar{z}\longrightarrow \bar{z}^\prime=\frac{\bar{a}\bar{z}+\bar{b}}{\bar{c}\bar{z}+\bar{d}},\label{5}
\end{eqnarray}
where $a,b,c,d\in\mathbb{C}$, $ad-bc=1$, $\alpha(z,\bar{z})$ is a generic scalar
function over $S^2$ and
$$K_\Lambda=\frac{1+\left|z\right|^2}{\left|az+b\right|^2+\left|cz+d\right|^2}.$$
The relations (\ref{3}), (\ref{4}) and (\ref{5}) form the so-called {\bf Bondi-Metzner-Sachs group} (BMS group)
and a 
direct inspection shows that it is the semidirect product between $SL(2,\mathbb{C})$, the universal cover 
of the proper ortochronous Lorentz group, and $N$, the set of scalar functions over $S^2$ (considered as a group
under addition). Furthermore there is an arbitrariness in the choice of the topology
of $N$, namely we have 
the freedom to impose a suitable regularity condition over each $\alpha:S^2\to\mathbb{R}$ i.e. $\alpha$
could lie in $C^\infty(S^2)$ (nuclear topology) or in $L^2(S^2)$ (Hilbert topology). There is no general
mathematical reason to prefer one of the two choices though, in a recent analysis \cite{Dappiaggi3}, it
appears that, from the point of view of an holographic correspondence, the 
nuclear topology is the unique one which allows to coherently project bulk
data to $\Im$.
Thus, we can conclude that the diffeomorphism group of the intrinsic and universal boundary structure of
any four dimensional asymptotically flat spacetime\footnote[2]{From the point of view of the AdS/CFT correspondence,
the boundary theory is invariant under the asymptotic symmetry group of an AdS spacetime. In 
the asymptotically flat scenario, the above introduced notion of diffeomorphism group of the boundary structure
coincides with the asymptotic symmetry group first derived in \cite{Bondi}.} is
$$BMS=SL(2,\mathbb{C})\ltimes C^\infty(S^2),$$
where $\ltimes$ stands for ``semidirect product'' and $C^\infty(S^2)$ is an
abelian group usually called ``supertranslations''.

\section{Kinematical and dynamical data for a field theory on null infinity}

The identification of $\Im$ as a candidate screen where to encode bulk data in an
asymptotically flat spacetime leads to the question whether it is possible to 
coherently define a (quantum) field theory living only on $\Im$ itself. By the light of our
previous discussion, it is natural to require the invariance of such a theory
under diffeomorphisms i.e. we need to construct a \emph{BMS
invariant field theory}. 

The first step in this quest consists on identifying the
set of possible free fields compatible with such a huge symmetry group and
theirs dynamic on $\Im$. This task can be completed following the pattern that
led Wigner to complete the same project for a Poincar\'e invariant theory living in
Minkowski spacetime i.e. a free field is a wave function transforming under a
unitary and irreducible representation (irrep.) of the Poincar\'e group. Thus, if we
stick to such a definition, substituting Poincar\'e with BMS invariance, we need
to classify and explicitly construct all the unitary irreps. of the
Bondi-Metner-Sachs group. This programme has been completed in the late
seventies by McCarthy both in the nuclear and in the Hilbert topology using
Mackey's theory of induced representations and we
will review here some important details (see \cite{Dappiaggi3, Arcioni} and
references therein for an exhaustive analysis).\\
Let us introduce the following concepts:
\begin{itemize}
\item the \emph{character} of $N=C^\infty(S^2)$ which is a continuous group
homomorphism $\chi:N\to U(1)$ which associates to each map $\chi$ a unique 
distribution $\beta$ lying in $N^*$, the topological dual
space of $N$, such that 
$$\chi(\alpha)=e^{i(\alpha,\beta)},\;\;\;\forall\alpha\in N$$
where $(\beta,\alpha)$ stands for the evaluation of the distribution $\beta$ 
with the suitable test function $\alpha$\footnote[3]{We drop from now on the
angular dependence of supertranslation restoring it if necessary to avoid
confusion.},
\item the \emph{orbit} of a character $\chi$ i.e. the set
$$\mathcal{O}_\chi:=\left\{g\chi\;|\;g=(\Lambda,\alpha)\in BMS\right\},$$
where $g\chi(\alpha)=\chi(g^{-1}\alpha)$ for any $\alpha\in N$.
\item the \emph{isotropy group} of $\chi$:
$$H_\chi:=\left\{g\in BMS\;|\; g\chi=\chi\right\}.$$
For the BMS group all the isotropy subgroups are $H_\chi=L_\chi\ltimes N$ where $L_\chi$ is
a closed subgroup of $SL(2,\mathbb{C})$ called \emph{little group}.
\end{itemize}

Thus, if we consider a character $\chi$ and a closed little group $L_\chi$, it
is possible to construct a unitary representation $U$ of $L_\chi\ltimes N$ acting on
a (non necessary finite dimensional) Hilbert space $\mathcal{H}$ as follows:
\begin{equation}\label{irrep.}
U(\Lambda,\alpha)\psi=\chi(\alpha)D(\Lambda)\psi,
\end{equation}
where $D(\Lambda)$ is a unitary representation of $L_\chi$.
The key step consists on showing that (\ref{irrep.}) induces a unitary and
irreducible representation of the BMS group which, thus, can be classified only
by means of the possible little groups\footnote[4]{This
statement is not throughout complete since a key achievement consists on establishing 
if the list of all the above irreducible representation is complete. We leave a
discussion of this problem to \cite{Dappiaggi3, Mc4}.} $L_\chi$. The latter
problem has been
studied in detail in \cite{Mc4} and the result consists on a plethora of subgroups of
$SL(2,\mathbb{C})$, the most notables being the connected ones i.e. $SU(2)$, 
$SO(2)$ and $\Delta$ the double cover of two dimensional Euclidean group. 
These groups represent the key ingredient to construct the kinematical
configurations of a BMS invariant free field theory; we are now entitled to
introduce the so-called \emph{induced wave function} which is a map transforming under a
unitary and irreducible representation of the full symmetry group i.e. a BMS
free field
\begin{equation}
\psi:\mathcal{O}_\chi=\frac{SL(2,\mathbb{C})}{L_\chi}\to\tilde{\mathcal{H}},
\end{equation}
where $\tilde{\mathcal{H}}$ is a suitable target Hilbert space and where $\psi$ 
transforms as
\begin{equation}\label{ind1}
\left(\Lambda\psi\right)(p)=\sqrt{\frac{d\mu(\Lambda
p)}{d\mu(p)}}D\left(\omega(p)^{-1}\Lambda\omega(\Lambda^{-1}p)\right)\psi(
\Lambda^{-1}p),\;\Lambda\in
SL(2,\mathbb{C})
\end{equation}
\begin{equation}\label{ind2}
\left(\alpha\psi\right)(p)=p(\alpha)\psi(p),\quad\alpha\in C^\infty(S^2)
\end{equation}
where $p$ is a generic point on the orbit 
$\mathcal{O}_\chi=SL(2,\mathbb{C})\chi$ and, thus, it is
a character. Furthermore $d\mu(p)$ is a suitably chosen measure on
$\mathcal{O}_\chi$, $\omega$ is a global section of the bundle
$\pi:SL(2,\mathbb{C})\to \frac{SL(2,\mathbb{C})}{L_\chi}$ and $D(\Lambda)$ a
unitary representation of $L_\chi$. 

The reader should notice that the induced wave function is defined in the space
of characters though, as stated previously, to each $\chi$ we can associate a
distribution over $S^2$ and consequently both (\ref{ind1}) and (\ref{ind2}) can
be equivalently read as maps over $N^*$. Furthermore it is possible to
associate to any of the above free fields a notion of mass with the following
argument. Consider $N=C^\infty(S^2)$ whose elements
can be seen as linear combinations of spherical harmonics $Y_{lm}(z,\bar{z})$. 
The set of the first four harmonics, i.e. $l=0,1$, span a four dimensional
$SL(2,\mathbb{C})$-invariant normal subgroup of $N$ isomorphic to $T^4$ (hence
the name ``supertranslations for $C^\infty(S^2)$).
Consider now the set of distributions $N^*$, the topological dual space
of $N$ and the annihilator $T^4_0=\left\{\beta\in
N^*\;|\;(\alpha,\beta)=0\;\forall\alpha\in T^4\subset C^\infty(S^2)\right\}$. It
is possible to introduce the projection \cite{Mc4}
$$\pi:N^*\to \left(T^4\right)^*\sim\frac{N^*}{\left(T^4\right)^0},$$
where $\sim$ stands for a non canonical isomorphism and where, to each 
distribution, it is associated an element in the space generated by the
dual spherical harmonics $Y^*_{lm}$ implicitly defined as $(Y^*_{lm},Y_{l^\prime
m^\prime})=\delta_{ll^\prime}\delta_{mm^\prime}$. Furthermore, since the above
projection is $SL(2,\mathbb{C})$-invariant and since, by construction, $(T^4)^*$
is isomorphic to $T^4$ it is possible to associate to each $\beta\in N^*$ a
scalar function over $S^2$ which is a linear combination of the first four
harmonics. The coefficients of such a combination identify a 4-vector (also
called the Poincar\'e momentum) which we indicate as $\pi[\beta]^\mu$ and
consequently we can introduce the quantity
$$m^2=\eta_{\mu\nu}\pi[\beta]^\mu\pi[\beta]^\nu,$$
which turns out to be a Casimir invariant (together with $sgn[\pi(\beta)_0]$) 
for the theory of \underline{faithful} unitary and irreducible representations of the BMS group \cite{Mc4}. 

Consequently, in analogy with the Poincar\'e counterpart, it is natural to 
identify $m^2$ as the squared mass of a BMS free field and $N^*$ as the space of
supermomenta\footnote[5]{This is completely equivalent to Wigner's construction for
a Poincar\'e invariant theory on Minkowski spacetime where $T^4$ played the role 
of coordinate space and $(T^4)^*$ that of momenta space.}. Eventually, according
to the analysis in \cite{Dappiaggi3, Mc4}, we have all the ingredients for a
full classification of the kinematical configurations of a BMS invariant free
field. In particular, discarding negative values of $m^2$, the results for the
connected little groups can be summarized in the following tabular:

\begin{center}
\begin{tabular}{|c|c|}
\hline
Orbit & possible value for the mass \\
\hline
$\frac{SL(2,\mathbb{C}}{SU(2)}$ & $m^2>0$ \\
\hline
$\frac{SL(2,\mathbb{C}}{SO(2)}$ & $m^2\geq 0$ \\
\hline
$\frac{SL(2,\mathbb{C}}{\Delta}$ & $m^2=0$\\
\hline
\end{tabular}
\end{center}

At this stage, we are far from completing Wigner programme since we have
also the chance to construct the dynamics of all the free fields only by means
of the theory of representations. The starting point consists on noticing that,
in physics, the concept of induced wave function is not commonly used whereas it
is widely introduced the so-called \emph{covariant wave function} i.e., in a BMS
language,
\begin{equation}
\psi^\prime:N^*\longrightarrow\mathcal{H}^\prime,
\end{equation}
where $\mathcal{H}^\prime$ is a suitable target Hilbert space and where
$\psi^\prime$ transforms under a unitary but not necessarily irreducible representation
of the BMS group i.e.
\begin{equation}\label{cov}
\left[U(g)\psi^\prime\right](\beta)=\chi_\beta(\alpha)\tilde{D}(\Lambda)\psi^\prime[\Lambda^{-1}\beta],\quad\forall
g=(\Lambda,\alpha)
\end{equation}
where $\tilde{D}(\Lambda)$ is a unitary $SL(2,\mathbb{C})$ representation and
$\chi_\beta$ stands for the unique character associated to the
$\beta$-distribution.

According to the previous discussion, it is immediate to realize that
(\ref{cov}) does not represent a free field since the irreducibility of the
representation is a request which cannot be easily given up. The rationale
behind Wigner argument is that the covariant wave function can be made
completely equivalent to (\ref{ind1}) and to (\ref{ind2}) if suitable
constraints are imposed. Beside some technical details, fully accounted in
\cite{Arcioni}, in a BMS framework, these constraints are threefold:
\begin{itemize}
\item an orbit equation which restricts the support of (\ref{cov}) from $N^*$ to
the finite dimensional orbit $\frac{SL(2,\mathbb{C})}{L_\chi}$:
$$[\beta-SL(2,\mathbb{C})\bar{\beta}]\psi^\prime(\beta)=0,$$
where $SL(2,\mathbb{C})\bar{\beta}$ stands for the action of
$SL(2,\mathbb{C})$\footnote[6]{For a generic element $\beta\in N^*$, $\Lambda\beta$ 
is defined in a distributional sense as
$\left(\Lambda\beta,\alpha\right)=(\beta,\Lambda^{-1}\alpha)$ for any $\alpha\in
N$.} on the distribution $\bar{\beta}$ chosen in such a way that $L_\chi\bar{\beta}=\bar{\beta}$.
\item a mass equation which associates to the $\psi^\prime$ a fixed value of
$m^2$ i.e.
$$[\eta_{\mu\nu}\pi(\beta)^\mu\pi(\beta)^\nu-m^2]\psi^\prime(\beta)=0$$
\item an equation which selects a unitary and irreducible representation
$D(\Lambda)$ of a 
little group $L_\chi$ inside $\tilde{D}(\Lambda)$ defined in (\ref{cov}). This can be achieved
introducing a suitable orthoprojector operator \cite{Arcioni, Barut} such that
$$\rho\left(\beta\right)\psi^\prime\left(\beta\right)=\psi^\prime\left(\beta\right).$$
\end{itemize}

The set of the above three equations represents in a momentum frame either the
constraints to impose on a covariant wave function to be equivalent to an induced wave
function either the equations of motion describing the dynamic of the associated
free field. As a matter of fact, the same construction leads in a Poincar\'e
invariant theory to the usual equations of Klein-Gordon, Dirac, Proca etc...
Hence, we have recovered in a BMS framework the full set of kinematical and
dynamical configurations for the free field theory and we summarize the results
by means of the explicit example of a BMS massive real scalar field i.e. the induced 
wave function \cite{Dappiaggi}:
$$\psi:\frac{SL(2,\mathbb{C})}{SU(2)}\longrightarrow\mathbb{R},$$
$$[g\psi](p)=p(\alpha)\psi(\Lambda^{-1}\beta).\quad g=(\Lambda,\alpha)$$
The above map is completely equivalent to its covariant counterpart
$$\psi^\prime:N^*\longrightarrow\mathbb{R},$$
\begin{equation}\label{KG1}
[U(\Lambda,\alpha)\psi^\prime](\beta)=e^{i(\alpha,\beta)}\psi^\prime(\Lambda^{-1}\beta),
\end{equation}
supplemented with the BMS Klein-Gordon equations of motion (the orthoprojector
$\rho\left(\beta\right)=1$)
\begin{equation}\label{KG2}
[\beta-\pi(\beta)]\psi^\prime\left(\beta\right)=0\quad
[\eta_{\mu\nu}\pi(\beta)^\mu\pi(\beta)^\nu-m^2]\psi^\prime(\beta)=0.
\end{equation}

\section{From boundary to bulk}
The key question we need to address is whether the above (intrinsic) boundary
data really encode the information from a fixed bulk theory. The main
obstruction within this respect lies in the universality of the null infinity
structure which is inherited by the BMS group i.e. a BMS field theory encodes a
priori the data from all the asymptotically flat spacetimes. 

In order to solve 
this problem, we face two candidate directions: the first consists on considering
a fixed bulk and on introducing a suitable projection of the physical data on
the boundary which are eventually interpreted as BMS invariant degrees of
freedom. This line of reasoning has been first discussed in \cite{Dappiaggi3}
and it will not be pursued here; conversely we will try to recover the bulk
information starting only from boundary data. Following \cite{Dappiaggi, Arcioni2,
 Dappiaggi2}, we start from the following remark: the physical
interpretation of Wigner construction strongly relies on a suitable
identification of the support of the covariant wave function with a submanifold
of the spacetime. In a Poincar\'e invariant theory, this is a straightforward
consequence either of the isomorphism between $T^4$ and $\mathbb{R}^4$ either of
the identification of $(T^4)^*$ with $T^4$ by means of the canonical pairing
between vectors and co-vectors induced by the metric $\eta_{\mu\nu}$.

In a BMS invariant theory, it is clearly not reasonable to look for an
identification of $N^*$ with $\Im$ and, instead, we follow a slightly different
path. It originates from an alternative formulation of general relativity
(nonetheless fully equivalent to Einstein's theory) which calls for dropping the
metric as the fundamental field variable. This approach, also known as
\emph{null surface formulation} of general relativity, has been developed in 
the mid nineties (see \cite{Frittelli2} and references therein) and the starting
point is a four dimensional asymptotically flat spacetime $M$, a generic point
$x^a\in M$ together with $L(x^a,x^{\prime a})=0$, the
light cone equation connecting $x_a$ to $x^\prime_a$. If we consider
$x^{\prime a}\in\Im^+$ and if we introduce the Bondi frame, then the boundary point
$x^{\prime a}$ can be parametrized as $(u,\theta,\varphi)$; furthermore the
equation $L(x^a,x^{\prime a})=0$ now depends on the $u$-variable and it can be 
inverted as
\begin{equation}\label{cut}
L\left(x^a,u,\theta,\varphi\right)=0\longrightarrow u=Z\left(x^a,\theta,\varphi\right).
\end{equation} 
The inverse of $L$ is not a priori unique and the set of $Z$ functions, also
known as ``cut functions'', represents the fundamental variable in the null
surface formulation of Einstein theory. Thus, within this respect, the metric
now becomes a functional dependent on the set of scalar maps over $M\times
S^2$ defined in (\ref{cut}); the full consistency of this approach and the
derivation of Einstein equations has been discussed in detail by several authors
(still refer to \cite{Frittelli2} and references therein) and we will not review
their results here. Conversely we concentrate on the interpretation of the
cut function namely, if we held fixed the boundary point on $\Im$, (\ref{cut})
represents the past light rays originating from $(u,\theta,\varphi)$. More
importantly, if we held fixed the bulk point, (\ref{cut}) describes the
intersection of the light cone originating from $x_a$ and intersecting $\Im$ on 
a $2$-surface which turns out to be homotopically equivalent to $S^2$.

Up to now we have discarded the chance that the light rays emanating from $x^a$
intersect at a certain point thus originating a caustic. Even if this pathology
should be cured in detail, to our purposes, it suffice to
notice, that it always exists a suitable neighbourhood of $\Im$ where $Z$ is a
unique and differentiable scalar function i.e. $Z_{x^a}(\theta,\varphi)\in
C^\infty(S^2)$. Consequently we may interpret each $Z_{x^a}$ as a BMS
supertranslation and, furthermore, these supertranslations reconstruct, up to
the conformal factor needed to compactify the asymptotically flat spacetime, the
full geometry of the bulk.

It is a natural suspicion that the reconstruction of the bulk data by means of
cut functions/supertranslations can be
transported from the geometrical setting to the field theoretical framework and
we conjecture that the information of a fixed asymptotically flat bulk spacetime
$M$ is encoded in the BMS fields whose support is compatible with the
supertranslations\footnote[7]{Though the support of the wave functions is
$N^*$ in a momentum frame, there is no contradiction in our hypothesis if we
remember that $N$ and $N^*$ form the Gelfand triplet $N\subset L^2(S^2)\subset
N^*$.} reconstructing $M$ in the null surface formalism.

Though we cannot supplement the above statement with a complete proof, we provide
here a concrete example reconstructing the 2-point function for the massive
scalar field in Minkowski spacetime starting from the BMS counterpart. Let us
thus start from (\ref{KG1}) and (\ref{KG2}) which are written in a
supermomentum frame and let us transform them in a supertranslation frame.
This task can be achieved by means of the infinite dimensional counterpart of
the Fourier transform discussed in \cite{Dappiaggi} and defined over the Hilbert
space of functions over a
second suitably chosen Hilbert space. Two ingredients are needed:
\begin{itemize}
\item since $N$ and $N^*$ form the Gelfand triplet $N\subset L^2(S^2)\subset
N^*$, we first need to notice that the support equation (\ref{KG2}) grants us
that the massive BMS scalar field lives on a mass hyperboloid generated by the
$SL(2,\mathbb{C})$ action over the $SU(2)$-fixed point in $N^*$. According to
the analysis in \cite{Mc4}, such a point is actually a smooth function over
$S^2$ and thus the entire orbit lies in $C^\infty(S^2)$ and consequently in
$L^2(S^2)$. Thus, if we require that each $\psi$, as in (\ref{KG1}), lies in the
space $\mathcal{H}^{\prime\prime}=L^2(L^2(S^2),\mu)$ where $\mu$ is the unique Gaussian measure associated to
$L^2(S^2)$, we are entitled to use the infinite dimensional Fourier transform 
theory.
\item we can define two important operators acting on an element $\psi$ lying in
$\mathcal{H}^{\prime\prime}$ namely the \emph{multiplication operator} along the
$\eta$-direction in $L^2(S^2)$:
$$Q_{\eta}\psi(x)=<\eta,x>\psi(x),$$
where $<,>$ is the internal product over $L^2(S^2)$ and the \emph{derivative}
operator along the $\eta$-direction:
$$\mathcal{D}_\eta\psi(x)=\lim\limits_{t\to 0}\frac{\psi(x+t\eta)-\psi(x)}{t}.$$
Both operators are related by means of the Fourier transform $\mathcal{F}$ as 
$\mathcal{F}Q_\eta=i\mathcal{D}_\eta\mathcal{F}$ and viceversa.
\end{itemize}
Bearing in mind these remarks, we can write (\ref{KG2}) in terms of multiplication
and derivative operators as
$$Q_{Y_{lm}}\psi(x)=0,\quad [\eta^{\mu\nu}Q_{e_\mu}Q_{e_\nu}-m^2]\psi(x)=0,$$
where $l>1$ and $e_\mu=\left\{Y_{00},...,Y_{11}\right\}$. Acting with the
Fourier transform, we end up with:
\begin{equation}
\mathcal{D}_{Y_{lm}}\psi(x)=0,\quad
[\eta^{\mu\nu}\mathcal{D}_{e_\mu}\mathcal{D}_{e_\nu}-m^2]\psi(x)=0.
\end{equation}
These can be read as the Euler-Lagrange equations for
\begin{equation}
L[\psi,\gamma_{lm}]=\psi(x)[\eta^{\mu\nu}\mathcal{D}_{e_\mu}\mathcal{D}_{e_\nu}-m^2]\psi(x)+\sum\limits_{l=0}^\infty\sum_{m=-l}^l\gamma_{lm}(x)\mathcal{D}_{Y_{lm}}\psi(x),
\end{equation}
where each $\gamma_{lm}(x)$ is a suitable Lagrange multiplier. We can now write
the action for a BMS Klein-Gordon field integrating the Lagrangian over the 
above introduced Gaussian measure $d\mu(x)$.
Eventually we can formulate a path-integral as
$$Z[\psi,\gamma_{lm}]=\int\limits_{\mathcal{C}}d\mu[\psi,\gamma_{lm}]
e^{iS[\psi,\gamma_{lm}]},\quad S[\psi,\gamma_{lm}]=\int\limits_{L^2(S^2)}
d\mu[x]L[\psi,\gamma_{lm}],$$
where $\mathcal{C}$ is a set of suitably chosen kinematical configurations. $Z$ 
reduces to the following expression \cite{Dappiaggi}:
$$Z[\psi,\gamma_{lm}]=\int\limits_{\mathcal{C}}d\mu[\psi]e^{i<\phi(x)B\phi(x)>},$$
where $<,>$ is the internal product on $L^2(S^2)$ and 
\begin{equation}
B=\eta^{\mu\nu}\mathcal{D}_{e_\mu}\mathcal{D}_{e_\nu}-m^2+\sum\limits_{l=2}^\infty\sum\limits_{m=-l}^l\frac{1}{2\zeta_{lm}}
(Q_{Y_{lm}}-\mathcal{D}_{Y_{lm}})\mathcal{D}_{Y_{lm}},
\end{equation}
where $\zeta_{lm}$ are arbitrary real number.
Whenever the partition function assumes the above expression, the Feynman
propagator $D_F(x_1-x_2)$ satisfies the equation
$$BD_F(x_1-x_2)=i\delta(x_1-x_2),$$
which, upon Fourier transform and using the definition of multiplication operator,
becomes the differential equation 
\begin{equation}\label{prop}
[\eta^{\mu\nu}k_\mu k_\nu-m^2+\sum\limits_{l=2}^\infty\sum\limits_{m=-l}^l\frac{1}{2\zeta_{lm}}
(k_{lm}-\mathcal{D}_{Y_{lm}})k_{lm}]D_F(k)=i,
\end{equation}
where $k_\mu=<e_\mu,k>$ and $k_{lm}=<Y_{lm},k>$. We can now appeal to our
conjecture and we look for those supertranslation/cut functions reconstructing
Minkowski spacetime. This is rather simple example from the point of view of the
null surface formulation since the light cone equation can be globally inverted
yielding $Z(x^\mu,\theta,\varphi)=x^\mu e_\mu$ where, as before, $e_\mu$ are only
the first four spherical harmonics. Similarly in a momentum frame we end up with
$Z(k^\mu,\theta,\varphi)=k^\mu e_\mu$ which, substituted in (\ref{prop}), gives
\begin{equation}\label{prop2}
[\eta^{\mu\nu}k_\mu k_\nu-m^2]D_F(k)=i\Longrightarrow
D_F(k)=\frac{i}{\eta^{\mu\nu}k_\mu k_\nu-m^2},
\end{equation}
which is exactly the Feynman propagator for a massive real scalar field in
Minkowski spacetime. 
\section{Conclusions}
We have discussed the kinematical and dynamical data of a free field theory 
living on the null
infinity boundary of an asymptotically flat spacetime and representing the
candidate to holographically encode the bulk information. 
We have also proposed a concrete way to reconstruct such information by means of
the null surface formulation of general relativity which allows us to interpret the BMS
supertranslations as the degrees of freedom encoding the geometrical
informations from the bulk. Starting from this rationale, we have recovered the
Feynman propagator for a Minkowski massive real scalar field using only the data from the
boundary counterpart. Though this is only a simple example, we believe that it
provides a good signal that holography can be implemented successfully on an
asymptotically flat manifold and that the BMS group should play a central role.
To conclude we wish to stress that the null nature of $\Im$ and the intrinsic
difficulty to work with an infinite dimensional group, such as the BMS,
naturally leads to treat holography in such a scenario with the rigorous
mathematical tools proper of the algebraic formulation of quantum field theory.
In particular we hope to provide rigorous holographic result similar to those
discussed by Rehren in \cite{Rehren, Rehren2} in the framework of AdS/CFT. A
preliminary analysis in \cite{Dappiaggi3} indicates that, at least in the
massless case, holography could be implemented in terms of Weyl algebra 
and the extension of these results to massive fields represents the next challenge.

\section*{Acknowledgments}
This work is supported by a grant from the Dipartimento di Fisica Nucleare e 
Teorica - Universit\'a di Pavia. The author is in debt with Valter Moretti and
Nicola Pinamonti for the long and fruitful discussions on  
algebraic quantum field theory and the role of the holographic principle within
this framework.

\end{document}